\documentclass[prd, preprint,nofootinbib,eqsecnum,superscriptaddress,12pt]{revtex4-1}
\pdfoutput=1

\newcount\showkey \showkey=0  

\usepackage{color}
\usepackage{tabularx}

\usepackage{hyperref}

\ifnum\showkey=1
\usepackage{seqsplit}
\usepackage[color]{showkeys}
\definecolor{refkey}{rgb}{0.9, 0.43, 0.63}
\definecolor{labelkey}{rgb}{0.59, 0.43, 0.63}
\usepackage{xstring}
\renewcommand*\showkeyslabelformat[1]{%
\noexpandarg%
\StrSubstitute{\(\{\)#1\(\}\)}{ }{\textvisiblespace}[\TEMP]%
\parbox[t]{\marginparwidth}{\raggedright\normalfont\small\ttfamily\expandafter\seqsplit\expandafter{\TEMP}}}
\fi

\usepackage{amsmath}
\usepackage{amssymb}
\usepackage{graphicx,multirow,subfigure}
\usepackage{float} 
\usepackage[T1]{fontenc}

\usepackage{rotating} 

\usepackage{enumitem}
\setlist[enumerate,2]{leftmargin=0.45em}

\usepackage{comment}

\usepackage[dvipsnames]{xcolor}

\usepackage[capitalize]{cleveref}    

\newcommand{\nn}{\nonumber}

\renewcommand{\bar}{\overline}

\usepackage{physics}
\usepackage{bbold}
\usepackage{mathtools}

\renewcommand \ket[1]{
        \left| #1 \right>
}

\renewcommand \bra[1]{
        \left< #1 \right|
}

\newcommand{\beq}{\begin{equation}}
\newcommand{\eeq}{\end{equation}}
\newcommand{\beqa}{\begin{eqnarray}}
\newcommand{\eeqa}{\end{eqnarray}}
\newcommand{\bea}{\begin{eqnarray}}
\newcommand{\eea}{\end{eqnarray}}
\newcommand{\bi}{\begin{itemize}}
\newcommand{\ei}{\end{itemize}}
\newcommand{\ben}{\begin{enumerate}}
\newcommand{\een}{\end{enumerate}}
\newcounter{mycount}
\newcommand{\pauseen}{\setcounter{mycount}{\value{enumi}}\end{enumerate}}
\newcommand{\resumeen}{\begin{enumerate}\setcounter{enumi}{\value{mycount}}}

\newcounter{defcount}
\newcounter{myenum}

\begin{document}

\title{A $U$-Spin Anomaly in Charm CP Violation}

\author{Stefan Schacht}
\email{stefan.schacht@manchester.ac.uk}
\affiliation{Department of Physics and Astronomy, University of Manchester, Manchester M13 9PL, United Kingdom}

\begin{abstract}
Recent LHCb data shows that the direct CP asymmetries of the decay modes $D^0\rightarrow \pi^+\pi^-$ and $D^0\rightarrow K^+K^-$ have the same sign, violating an improved $U$-spin limit sum rule in an unexpected way at $2.1\sigma$. From the new data, we determine for the first time the imaginary part of the CKM-subleading, $U$-spin breaking $\Delta U=1$ correction to the $U$-spin limit $\Delta U=0$ amplitude. The imaginary part of the $\Delta U=0$ amplitude is determined by $\Delta a_{CP}^{\mathrm{dir}}$. The corresponding strong phases are yet unknown and could be extracted in the future from  time-dependent measurements. Assuming $\mathcal{O}(1)$ strong phases due to non-perturbative rescattering, we find the ratio of $U$-spin breaking to $U$-spin limit contributions to the CKM-subleading amplitudes to be $(173^{+85}_{-74})\%$. This highly exceeds the Standard Model (SM) expectation of $\sim 30\%$ $U$-spin breaking, with a significance of $1.95\sigma$. If this puzzle is confirmed with more data in the future, in the SM it would imply the breakdown of the $U$-spin expansion in CKM-subleading amplitudes of charm decays. The other solution are new physics models that generate an additional $\Delta U=1$ operator, leaving the $U$-spin power expansion intact. Examples for the latter option are an extended scalar sector or flavorful $Z'$ models.
\end{abstract}

\maketitle


\section{Introduction \label{sec:introduction}}

After the discovery of CP violation in charm decays \cite{Aaij:2019kcg}, recently, there have again been several important advances in the measurement of mixing and CP violation in charm decays~\cite{Maccolini:2022, Belle:2019xha, LHCb:2021rou, LHCb:2021rdn, LHCb:2021vmn, LHCb:2021ykz, LHCb:2021dcr,  LHCb:2022gnc}, see Ref.~\cite{HFLAV:2022pwe} for most recent world averages and global fits.  
Also theoretically, charm CP violation obtains a 
lot of attention right now~\cite{Grossman:2019xcj, Chala:2019fdb, Li:2019hho, Soni:2019xko, Cheng:2019ggx, Buccella:2019kpn, Dery:2019ysp, Calibbi:2019bay, Dery:2021mll, Bause:2020obd, Kagan:2020vri, Buras:2021rdg, Acaroglu:2021qae, Schacht:2021jaz, Gavrilova:2022hbx}, 
see earlier Refs.~\cite{Einhorn:1975fw, Abbott:1979fw, Buras:1985xv, Golden:1989qx, Buccella:1992sg, Falk:2001hx, Brod:2012ud, Bhattacharya:2012ah, Franco:2012ck, Hiller:2012xm, Nierste:2017cua, Nierste:2015zra, Muller:2015lua, Muller:2015rna, Grossman:2018ptn, Buccella:1994nf, Grossman:2006jg, Artuso:2008vf, Khodjamirian:2017zdu, Buccella:2013tya, Cheng:2012wr, Feldmann:2012js, Li:2012cfa, Atwood:2012ac, Grossman:2012ry, Yu:2017oky, Brod:2011re, Pirtskhalava:2011va, Altmannshofer:2012ur, Grossman:2012eb}.
The most recent news is the first evidence of a non-vanishing CP asymmetry 
\begin{align}
a_{CP}^{\mathrm{dir}}(f) &\equiv \frac{
	\vert \mathcal{A}(D^0\rightarrow f)\vert^2 - \vert \mathcal{A}(\overline{D}^0\rightarrow f) \vert^2
}{
	\vert \mathcal{A}(D^0\rightarrow f)\vert^2 + \vert \mathcal{A}(\overline{D}^0\rightarrow f) \vert^2
}  
\end{align}
in a single decay~\cite{Maccolini:2022}, namely $D^0\rightarrow \pi^+\pi^-$. The knowledge of both CP asymmetries~\cite{Maccolini:2022}
\begin{align}
a_{CP}^{\mathrm{dir}}(D^0\rightarrow K^+K^-) &= (7.7\pm 5.7)\cdot 10^{-4} \,, \\
a_{CP}^{\mathrm{dir}}(D^0\rightarrow \pi^+\pi^-) &=  (23.2\pm 6.1)\cdot 10^{-4}
\end{align}
gives important advantages compared to the combination 
\begin{align}
\Delta a_{CP}^{\mathrm{dir}} &\equiv a_{CP}^{\mathrm{dir}}(D^0\rightarrow K^+K^-) - a_{CP}^{\mathrm{dir}}(D^0\rightarrow \pi^+\pi^-) 
\end{align}
only. The reason is that a separate measurement of both CP asymmetries allows to test the $U$-spin expansion in the amplitude contributions which are relatively suppressed by Cabibbo-Kobayashi-Maskawa (CKM) matrix elements compared to the leading singly-Cabibbo-suppressed (SCS) amplitude.
In fact, we can now probe the $U$-spin limit sum rule for the sum of CP asymmetries~\cite{Grossman:2006jg, Pirtskhalava:2011va, Hiller:2012xm, Grossman:2012ry} 
\begin{align}
\Sigma a_{CP}^{\mathrm{dir}} &\equiv a_{CP}^{\mathrm{dir}}(D^0\rightarrow K^+K^-) + a_{CP}^{\mathrm{dir}}(D^0\rightarrow \pi^+\pi^-) \overset{\text{$U$-spin limit}}{=} 0\,, \label{eq:sigma-sum-rule} 
\end{align}
which is violated at $2.7\sigma$~\cite{Maccolini:2022}.
Remarkably, Eq.~(\ref{eq:sigma-sum-rule}) predicts that $a_{CP}^{\mathrm{dir}}(D^0\rightarrow K^+K^-) $ and $a_{CP}^{\mathrm{dir}}(D^0\rightarrow \pi^+\pi^-)$ have opposite signs, but in fact the measurement shows that they have the same sign. 

An improved version of the sum rule Eq.~(\ref{eq:sigma-sum-rule}) is given as~\cite{Grossman:2006jg, Pirtskhalava:2011va, Hiller:2012xm, Grossman:2012ry}
\begin{align}
\frac{\Gamma(D^0\rightarrow K^+K^-)}{\Gamma(D^0\rightarrow \pi^+\pi^-)} &\overset{\text{$U$-spin limit}}{=} -\frac{a_{CP}^{\mathrm{dir}}(D^0\rightarrow \pi^+\pi^-)}{ a_{CP}^{\mathrm{dir}}(D^0\rightarrow K^+K^-)}\,. \label{eq:improved-sigma-sum-rule} 
\end{align}
The sum rules Eqs.~(\ref{eq:sigma-sum-rule}, \ref{eq:improved-sigma-sum-rule}) belong to a category of $U$-spin sum rules which are based on the complete interchange of $s$ and $d$ quarks~\cite{Gronau:2000zy, Fleischer:1999pa, Gronau:2000md}.
Inserting the experimental measurements listed in Table~\ref{tab:exp-input-data} below, we obtain 
\begin{align}
\frac{\Gamma(D^0\rightarrow K^+K^-)}{\Gamma(D^0\rightarrow \pi^+\pi^-)} &=  2.81 \pm 0.06
\end{align}
and 
\begin{align}
-\frac{a_{CP}^{\mathrm{dir}}(D^0\rightarrow \pi^+\pi^-)}{ a_{CP}^{\mathrm{dir}}(D^0\rightarrow K^+K^-)}  &=  -3.01^{+0.95}_{-5.95}\,,
\end{align}
\emph{i.e.} altogether
\begin{align}
- \frac{\Gamma(D^0\rightarrow K^+K^-)}{\Gamma(D^0\rightarrow \pi^+\pi^-)} \frac{ a_{CP}^{\mathrm{dir}}(D^0\rightarrow K^+K^-)}{ a_{CP}^{\mathrm{dir}}(D^0\rightarrow \pi^+\pi^-)} &= -0.93^{+0.62}_{-0.41} \neq +1. \label{eq:sum-rule-deviation}
\end{align} 
The improved $U$-spin limit sum rule Eq.~(\ref{eq:improved-sigma-sum-rule}) is broken at $2.1\sigma$, because Eq.~(\ref{eq:sum-rule-deviation}) has the \lq\lq{}wrong\rq\rq{} sign. While $U$-spin breaking is expected, because $U$-spin is only an approximate symmetry of QCD, 
the amount of breaking goes beyond the Standard Model expectations of $\varepsilon\sim m_s/\Lambda_{\mathrm{QCD}}\sim 30\%$ at $1.9\sigma$. 

In this article, we analyze the implications of the new charm CP measurements in more detail, extracting the CKM-subleading $\Delta U=1$ contributions to the amplitudes of $D^0\rightarrow K^+K^-$ and $D^0\rightarrow \pi^+\pi^-$ decays. In the SM these are generated from the tensor product of the $U$-spin limit $\Delta U=0$ operator with the $U$-spin breaking triplet operator~\cite{Jung:2009pb, Gavrilova:2022hbx}. 

After briefly reviewing the application of SU(3)$_F$ methods in charm decays in Sec.~\ref{sec:review}, we summarize our notation in Sec.~\ref{sec:notation}. In Sec.~\ref{sec:analytic} we recapitulate how to completely solve the system of two-body $D^0$ decays to kaons and pions. We also show explicitly how to extract in principle the strong phases of the CKM-subleading $\Delta U=1$ and $\Delta U=0$ hadronic matrix elements from time-dependent CP violation. In Sec.~\ref{sec:numerics} we present our numerical results. Finally, in Sec.~\ref{sec:predictions} we give predictions and options for interpretations in terms of new physics models that can be tested with future and more precise data. We conclude in Sec.~\ref{sec:conclusions}.

\section{Review of SU(3)$_F$-breaking in Charm Decays \label{sec:review}}

The application of SU(3)$_F$ methods in particle physics have their roots in spectroscopy, namely the \lq\lq{}eightfold way\rq\rq{} for the description of the spectrum of the meson and baryon octets~\cite{Gell-Mann:1961omu, Neeman:1961jhl}. In spectroscopy, SU(3)$_F$ has proven to be an extremely useful ordering principle. For example, SU(3)$_F$-limit predictions agree with the baryon octet mass splitting with an accuracy of $10\%$~\cite{Greiner:1989eu}. 
Furthermore, the Gell-Mann--Okubo mass formula \cite{Gell-Mann:1961omu, Okubo:1961jc} demonstrated that by including SU(3)$_F$-breaking effects in a systematic way the precision of predictions can be significantly improved. 
We know therefore that SU(3)$_F$ is a very trustable technique for the particle spectrum.
The question is if the same applies also to decay rates, in particular for charm decays.

The nominal size of SU(3)$_F$-breaking for decay amplitudes can be estimated from the ratio of 
the decay constants~\cite{FlavourLatticeAveragingGroupFLAG:2021npn}
\begin{align}
\frac{f_K}{f_{\pi}} - 1 \sim 0.2\,. 
\end{align}
Now, two important examples where it looks naively as if $U$-spin is broken by 
$\mathcal{O}(1)$ are given in terms of the ratios 
\begin{align}
\frac{\mathcal{B}(D^0\rightarrow K^+K^-)}{\mathcal{B}(D^0\rightarrow \pi^+\pi^-)} &\sim 3\,,  \label{eq:SU3-breaking-KKpipi}\\
\frac{\mathcal{B}(D^0\rightarrow K_S K_S)}{\mathcal{B}(D^0\rightarrow K^+K^-)} &\sim 0.03\,. \label{eq:SU3-breaking-KSKS}
\end{align}
In the strict SU(3)$_F$ limit, neglecting also differences from phase space effects, we have:
\begin{align}
\frac{\mathcal{B}(D^0\rightarrow K^+K^-)}{\mathcal{B}(D^0\rightarrow \pi^+\pi^-)} &= 1\,,\\
\frac{\mathcal{B}(D^0\rightarrow K_S K_S)}{\mathcal{B}(D^0\rightarrow K^+K^-)} &= 0\,,
\end{align}
in clear contradiction with the experimental measurements Eqs.~(\ref{eq:SU3-breaking-KKpipi}) and (\ref{eq:SU3-breaking-KSKS}).
However, already in Ref.~\cite{Savage:1991wu} it was realized that Eq.~(\ref{eq:SU3-breaking-KKpipi}) can actually be explained by $\varepsilon\sim 30\%$ SU(3)$_F$-breaking on the amplitude. 
This can be understood as follows: Already for $\varepsilon\sim 30\%$, very roughly the ratio of branching ratios can be estimated as 
\begin{align}
\frac{\left(1+\varepsilon\right)^2}{\left(1-\varepsilon\right)^2} &\sim 3\,. \label{eq:rough-estimate}
\end{align}
That means, Eq.~(\ref{eq:SU3-breaking-KKpipi}) can be consistently explained with SU(3)$_F$ breaking of $\sim 30\%$ on the amplitude.
Note that the rough illustration Eq.~(\ref{eq:rough-estimate}) also demonstrates that higher order contributions may be important, as at linear order the left-hand side of Eq.~(\ref{eq:rough-estimate}) results in $\sim 2$, and only at $\mathcal{O}(\varepsilon^2)$ it reaches $\sim 3$. Below, we extract first and second order $U$-spin breaking from branching ratio data, see Eqs.~(\ref{eq:param-1})--(\ref{eq:param-4}) and Table~\ref{tab:results}. For the CKM-leading amplitudes, which dominate the branching ratios, our results support that $\varepsilon\sim 0.3$ and $\varepsilon^2\sim 0.1$, consistent with the $U$-spin power counting.

Coming now to the second example, as $\mathcal{B}(D\rightarrow K_SK_S)$ vanishes in the SU(3)$_F$ limit, we can estimate the corresponding amplitude-level SU(3)$_F$ breaking roughly as
\begin{align}
\varepsilon' &\sim \sqrt{\frac{\mathcal{B}(D^0\rightarrow K^0 \overline{K}^0)}{\mathcal{B}(D^0\rightarrow K^+K^-)}} 
	= \sqrt{\frac{2 \mathcal{B}(D^0\rightarrow K_S K_S)}{\mathcal{B}(D^0\rightarrow K^+K^-)}} \sim 0.26\,, \label{eq:KSKS-KpKm-ratio}
\end{align}
again consistent with the nominal size of SU(3)$_F$ breaking.
Here we use the experimental values~\cite{Workman:2022ynf} 
\begin{align}
\mathcal{B}(D^0\rightarrow K_S K_S) &= (1.41\pm 0.05)\cdot 10^{-4}\,, \\
\mathcal{B}(D^0\rightarrow K^+K^-)  &= (4.08\pm 0.06)\cdot 10^{-3}\,, 
\end{align}
and, due to Bose symmetry, see e.g.~Ref.~\cite{Nierste:2015zra}
\begin{align}
\mathcal{A}(D^0\rightarrow K_SK_S ) &= -\frac{1}{\sqrt{2}} \mathcal{A}(D^0\rightarrow \overline{K}^0 K^0)\,.
\end{align}
It follows that Eqs.~(\ref{eq:SU3-breaking-KKpipi}, \ref{eq:SU3-breaking-KSKS}) can not be used as an argument that SU(3)$_F$ is broken at $\mathcal{O}(1)$ for charm decays. 

We note that if one would adopt additional theory assumptions in terms of a $1/N_c$ power counting~\cite{tHooft:1973alw, Buras:1985xv} on top of the SU(3)$_F$ expansion,
in the $1/N_c$ limit one can factorize the tree amplitude of non-leptonic charm decays, see e.g. Ref.~\cite{Muller:2015lua}.
However, in this case the factorizable $U$-spin breaking of the tree amplitudes alone does not suffice in order to explain the SU(3)$_F$ breaking in Eq.~(\ref{eq:SU3-breaking-KKpipi})~\cite{Cheng:2010ry}.
In the topological diagram approach, besides the tree amplitude $T$, the branching ratios of  
 $D^0\rightarrow K^+K^-$ and $D^0\rightarrow \pi^+\pi^-$ depend also on exchange diagrams $E$ and SU(3)$_F$-breaking combinations of penguin contractions of the tree operator $P_{\mathrm{break}}$, see the parametrizations in Refs.~\cite{Brod:2012ud, Muller:2015lua}. Therefore, under the assumption of a $1/N_c$ power counting, in order to explain Eq.~(\ref{eq:SU3-breaking-KKpipi}), additional contributions to the SU(3)$_F$-breaking have to come from these contributions. 
At first glance this seems counterintuitive, as $E$ and $P_{\mathrm{break}}$ are formally $1/N_c$ suppressed relative to $T$, which would also affect the possible amount of SU(3)$_F$ breaking. 
However, there are two contributions to these respective topological diagrams, which are both 
suppressed by $1/N_c$, and which stem from the Hamiltonian~\cite{Muller:2015lua}
\begin{align}
H_W 	&\propto C_1 Q_1 + C_2 Q_2\,, \label{eq:Hamiltonian}
\end{align}
where
\begin{align}
C_1 &\sim \mathcal{O}(1/N_c)\,, &
C_2 &\sim \mathcal{O}(1)\,, \\
\bra{PP'}Q_1\ket{D} &\sim \mathcal{O}(1)\,, & 
\bra{PP'}Q_2\ket{D} &\sim \mathcal{O}(1/N_c)\,. \label{eq:1Nc-matrix-elements} 
\end{align}
A priori, it is unclear how the two terms of order $\mathcal{O}(1/N_c)$ from Eq.~(\ref{eq:Hamiltonian}) interfere. This depends on the assumptions one makes about the respective matrix elements and can at this time not be 
determined from first principles.
The fit in Ref.~\cite{Muller:2015lua} shows the existence of a solution that is compatible with $1/N_c$ counting, namely when both contributions interfere constructively, see Fig.~3(c) therein.
This leads then to a large $E/T$ ratio, see also Refs.~\cite{Bhattacharya:2009ps, Cheng:2010ry, Brod:2011re}.

The fit result can be understood already when considering the single decay mode $\mathcal{B}(D \rightarrow K_SK_S)$, which only depends on SU(3)$_F$-breaking exchange diagrams. Eq.~(\ref{eq:KSKS-KpKm-ratio}) determines their relative size as $\sim 0.26$. 
In case of a constructive interference of the matrix elements of Eq.~(\ref{eq:Hamiltonian}), together with the estimates Eq.~(\ref{eq:1Nc-matrix-elements}), we obtain the rough estimate 
\begin{align}
\varepsilon  ( \vert C_1\vert + \vert C_2/N_c\vert ) \sim 0.24\,, \label{eq:KSKS-estimate}
\end{align}  
where we use $C_2=1.2$ and $C_1=-0.4$~\cite{Muller:2015lua}. The estimate Eq.~(\ref{eq:KSKS-estimate}) reproduces the measurement Eq.~(\ref{eq:KSKS-KpKm-ratio}) up to 8\%. 
At the same time, the fit in Ref.~\cite{Muller:2015lua} finds that the large SU(3)$_F$-breaking exchange diagrams together with the broken penguin can also explain Eq.~(\ref{eq:SU3-breaking-KKpipi}).

Global fits in the pure group-theoretical approach~\cite{Hiller:2012xm} agree with the approach employing topological diagrams~\cite{Muller:2015lua} in that the maximal needed linear SU(3)$_F$ breaking in the CKM-leading amplitudes is given as $\varepsilon\sim 30\%$.

We can test the SU(3)$_F$ expansion also beyond linear order breaking effects.
For the ratio 
\begin{align}
R_{DPP} &\equiv \frac{\vert \mathcal{A}(D^0\rightarrow K^+K^-)/(V_{cs} V_{us})\vert + 
	\vert \mathcal{A}(D^0\rightarrow \pi^+\pi^-)/(V_{cd} V_{ud})\vert
		}{
	\vert \mathcal{A}(D^0\rightarrow K^+\pi^-)/(V_{cd} V_{us})\vert + 
		\vert \mathcal{A}(D^0\rightarrow K^-\pi^+)/(V_{cs} V_{ud}) \vert
		} - 1  
\end{align}
the SU(3)$_F$ expansion predicts that it is proportional to second order SU(3)$_F$-breaking effects~\cite{Brod:2012ud, Grossman:2012ry, Falk:2001hx}
\begin{align}
R_{DPP}^{\mathrm{th}} &= \mathcal{O}(\varepsilon^2)\,. \label{eq:U-spin-second-order-th}
\end{align}
The experimental branching ratio measurements give
\begin{align}
R_{DPP}^{\mathrm{exp}} &= 0.046 \pm 0.008\,, \label{eq:U-spin-second-order}    
\end{align}
confirming the theory prediction Eq.~(\ref{eq:U-spin-second-order-th}).
If $U$-spin breaking were $\mathcal{O}(1)$, the second order $U$-spin breaking contributions that are isolated in Eq.~(\ref{eq:U-spin-second-order}) would still be $\mathcal{O}(1)$. Instead, it is, as expected, consistent with $\mathcal{O}(\varepsilon^2)$.

Although we see many examples where the SU(3)$_F$ expansion, including $U$-spin, is applied with great success, it is still an open question how trustable it is in general. 
Therefore, we seek to test the validity of $U$-spin at every possible opportunity. 

Below, from recent data, we identify a new puzzle that appears in the CKM-subleading amplitude contributions to charm decays, as opposed to the CKM-leading contributions discussed above.

\section{Notation \label{sec:notation}}

We use the notation of Ref.~\cite{Grossman:2019xcj} which we shortly summarize in this section.
In the SM, the Hamiltonian of SCS charm decays has the $U$-spin structure
\begin{align}
\mathcal{H}_{\mathrm{eff}} \sim \Sigma (1,0) - \frac{\lambda_b}{2} (0,0)\,,
\end{align} 
with $(i,j)=\mathcal{O}^{\Delta U=i}_{\Delta U_3=j}$ and the CKM matrix element combinations
\begin{align}
\Sigma &\equiv \frac{V_{cs}^*V_{us} - V_{cd}^* V_{ud}}{2}\,, &
-\frac{\lambda_b}{2} &\equiv -\frac{V_{cb}^* V_{ub}}{2} = \frac{V_{cs}^*V_{us} + V_{cd}^* V_{ud}}{2}\,.
\end{align}
Amplitudes of SCS charm decays can then be written as 
\begin{align}
\mathcal{A} &= \Sigma A_{\Sigma} - \frac{\lambda_b}{2} A_b\,. \label{eq:SCS-decays}
\end{align}
We use the following parametrization of $U$-spin related two-body $D^0$ decays to kaons and pions~\cite{Grossman:2019xcj, Brod:2012ud} 
\begin{align}
\mathcal{A}(K\pi) &= \mathcal{A}(\overline{D}^0\rightarrow K^+\pi^-) = V_{cs} V_{ud}^*\left(t_0 - \frac{1}{2} t_1 \right)\,,    & \text{(CF)}  & \label{eq:param-1}\\
\mathcal{A}(\pi\pi) &= \mathcal{A}(\overline{D}^0\rightarrow \pi^+\pi^-) = -\Sigma^* \left(t_0 + s_1 + \frac{1}{2} t_2\right) - \lambda_b^* \left( p_0 - \frac{1}{2} p_1 \right)\,,  & \text{(SCS)} &  \label{eq:param-2}\\
\mathcal{A}(KK) &= \mathcal{A}(\overline{D}^0\rightarrow K^+K^-) = \Sigma^*\left(t_0 - s_1 + \frac{1}{2} t_2 \right) - \lambda_b^* \left( p_0 + \frac{1}{2} p_1\right)\,, & \text{(SCS)} &  \label{eq:param-3}\\
\mathcal{A}(\pi K) &= \mathcal{A}(\overline{D}^0\rightarrow \pi^+K^- ) = V_{cd} V_{us}^* \left(t_0 + \frac{1}{2} t_1 \right)\,,  & \text{(DCS)} & \label{eq:param-4} 
\end{align}
where the decays are classified according to their suppression with Wolfenstein-$\lambda$ as Cabibbo-favored (CF), SCS and doubly-Cabibbo suppressed (DCS). 
The subscripts of the parameters in Eqs.~(\ref{eq:param-1})--(\ref{eq:param-4}) indicate the corresponding order in the $U$-spin expansion.
We employ the normalized parameters
\begin{align}
\tilde{t}_1 \equiv \frac{t_1}{t_0}\,,\qquad 
\tilde{t}_2 \equiv \frac{t_2}{t_0}\,,\qquad
\tilde{s}_1 \equiv \frac{s_1}{t_0}\,,\qquad
\tilde{p}_0 \equiv \frac{p_0}{t_0}\,,\qquad
\tilde{p}_1 \equiv \frac{p_1}{t_0}\,. \label{eq:normalized-parameters}
\end{align} 
The amplitudes are normalized such that
\begin{align}
\mathcal{B}(D\rightarrow PP') &= \vert \mathcal{A}\vert^2 \cdot \mathcal{P}(D,P,P')\,,\\
\mathcal{P}(D,P,P') &= \frac{\tau_D}{16 \pi m_D^3} \sqrt{
	\left(m_D^2 - (m_P - m_{P'} )^2 \right) 
	\left(m_D^2 - (m_P + m_{P'} )^2 \right)
	}\,,
\end{align} 
and \cite{Golden:1989qx, Nierste:2017cua, Pirtskhalava:2011va}
\begin{align}
a_{CP}^{\mathrm{dir}} &= \mathrm{Im}\left(\frac{\lambda_b}{\Sigma}\right) \mathrm{Im}\left(\frac{A_b}{A_{\Sigma}}\right)\,.
\end{align}
Furthermore, we write the amplitudes without CKM factors as $A(f)$ for CF and DCS decays and $A(f)\equiv A_{\Sigma}(f)$, $A_{b}(f)$ for SCS decays. 

Following Ref.~\cite{Grossman:2019xcj} we also use the observable combinations
\begin{align}
R_{K\pi} &\equiv \frac{ 
	\vert A(K\pi)\vert^2 - \vert A(\pi K)\vert^2
	}{ 
	\vert A(K\pi)\vert^2 + \vert A(\pi K)\vert^2
	}\,, \\
R_{KK,\pi\pi} &\equiv \frac{
	\vert A(KK)\vert^2 - \vert A(\pi \pi)\vert^2
	}{
	\vert A(KK)\vert^2 + \vert A(\pi \pi)\vert^2
	}\,,\\
R_{KK,\pi\pi, K\pi} &\equiv \frac{
	\vert A(KK)\vert^2 + \vert A(\pi\pi)\vert^2 - \vert A(K\pi)\vert^2 - \vert A(\pi K)\vert^2
	}{
	\vert A(KK)\vert^2 + \vert A(\pi\pi)\vert^2 + \vert A(K\pi)\vert^2 + \vert A(\pi K)\vert^2
	}\,.
\end{align}
The strong phase between CF and DCS $D^0$ decays is defined as 
\begin{align}
\delta_{K\pi} \equiv \mathrm{arg}\left(\frac{\mathcal{A}( D^0\rightarrow K^+\pi^-) }{\mathcal{A}( D^0\rightarrow K^-\pi^+ ) } \right)\,.
\end{align}
For convenience, we define the strong phases $\delta_{KK}$ and $\delta_{\pi\pi}$ slightly different from Ref.~\cite{Grossman:2019xcj} as
\begin{align}
\delta_{KK} &\equiv\mathrm{arg}\left(\frac{A_b(KK)}{A_{\Sigma}(KK)}\right)\,, & 
\delta_{\pi\pi} &\equiv \mathrm{arg}\left(\frac{A_b(\pi\pi)}{A_{\Sigma}(\pi\pi)}\right)\,.
\end{align}

\section{Solving for underlying theory parameters \label{sec:analytic}}  

In the convention of Ref.~\cite{Grossman:2019xcj}, the parametrization Eqs.~(\ref{eq:param-1})--(\ref{eq:param-4}) has the following eight real parameters, not counting the normalization $t_0$: 
\begin{align}
\mathrm{Re}(\tilde{t}_1), \quad \mathrm{Im}(\tilde{t}_1),\quad
\tilde{t}_2,\quad
\tilde{s}_1,\quad
\mathrm{Re}(\tilde{p}_0), \quad \mathrm{Im}(\tilde{p}_0),\quad
\mathrm{Re}(\tilde{p}_1), \quad\mathrm{Im}(\tilde{p}_1).
\end{align}
We can solve the complete system to order $\mathcal{O}(\varepsilon^2)$ as follows~\cite{Grossman:2019xcj}
\begin{align}
\mathrm{Re}(\tilde{t}_1) &= -R_{K\pi}\,, \label{eq:solution-1}\\
\mathrm{Im}(\tilde{t}_1) &= -\tan\left(\delta_{K\pi}\right)\,, \label{eq:solution-2}\\
\tilde{t}_2 &= 2 R_{KK,\pi\pi,K\pi} - \frac{1}{4} R_{KK,\pi\pi}^2 + \frac{1}{4} R_{K\pi}^2 + \frac{1}{4} \tan^2(\delta_{K\pi})\,, \label{eq:solution-3}\\
\tilde{s}_1 &= -\frac{1}{2} R_{KK,\pi\pi}\,, \label{eq:solution-4}\\
\mathrm{Im}(\tilde{p}_0) &= \frac{1}{4\, \mathrm{Im}(\lambda_b/\Sigma)} \Delta a_{CP}^{\mathrm{dir}}\,, \label{eq:solution-5}\\
\mathrm{Im}(\tilde{p}_1) &= \frac{1}{2\, \mathrm{Im}(\lambda_b/\Sigma)}\left( \Sigma a_{CP}^{\mathrm{dir}} + \frac{1}{2} R_{KK,\pi\pi} \Delta a_{CP}^{\mathrm{dir}} \right)\,, \label{eq:solution-6}\\ 
\mathrm{Re}(\tilde{p}_0) &= \frac{1}{4}\left( \mathrm{Re}\left( \frac{A_b(D^0\rightarrow K^+K^-) }{ A_{\Sigma}(D^0\rightarrow K^+K^-)}\right) - \mathrm{Re}\left( \frac{A_b(D^0\rightarrow \pi^+\pi^- ) }{A_{\Sigma}(D^0\rightarrow \pi^+\pi^-) }  \right) \right)\,, \label{eq:solution-7}\\
\mathrm{Re}(\tilde{p}_1) &=  \frac{1}{2}\left( \mathrm{Re}\left( \frac{A_b(D^0\rightarrow K^+K^-     ) }{ A_{\Sigma}(D^0\rightarrow K^+K^-)   }\right) + 
			     \mathrm{Re}\left( \frac{A_b(D^0\rightarrow \pi^+\pi^- ) }{ A_{\Sigma}(D^0\rightarrow \pi^+\pi^-) }\right)\right) \nn\\
&\quad	+ \frac{1}{4} R_{KK,\pi\pi} \left( \mathrm{Re}\left( \frac{A_b(D^0\rightarrow K^+K^-) }{ A_{\Sigma}(D^0\rightarrow K^+K^-)}\right) - \mathrm{Re}\left( \frac{A_b(D^0\rightarrow \pi^+\pi^- ) }{A_{\Sigma}(D^0\rightarrow \pi^+\pi^-) }  \right)\right)\,. \label{eq:solution-8}
\end{align}
Note that $\tan\delta_{K\pi} \approx \delta_{K\pi}$. 
Furthermore, we have to $\mathcal{O}(\varepsilon^2)$: 
\begin{align}
\frac{1/2\, \mathrm{Im}(\tilde{p}_1)}{ \mathrm{Im}(\tilde{p}_0) }  = \frac{ \Sigma a_{CP}^{\mathrm{dir}}}{ \Delta a_{CP}^{\mathrm{dir}} } + \frac{1}{2} R_{KK,\pi\pi}\,.  \label{eq:observable-relation}
\end{align}

The parameters $\mathrm{Re}(\tilde{p}_0)$ and $\mathrm{Re}(\tilde{p}_1)$ can be determined from time-dependent measurements.
In the following, we write the equations for $\mathrm{Re}(\tilde{p}_0)$ and $\mathrm{Re}(\tilde{p}_1)$ in a more convenient form in terms of the phases 
$\cot\delta_{KK}$ and $\cot\delta_{\pi\pi}$. 
These are related to the parametrization Eqs.~(\ref{eq:param-1})--(\ref{eq:param-4}) as 
\begin{align}
\cot\delta_{KK} &= \frac{\mathrm{Re}(A_b(KK)/A_{\Sigma}(KK))}{\mathrm{Im}(A_b(KK)/A_{\Sigma}(KK))}\,,  &
\cot\delta_{\pi\pi} &= \frac{\mathrm{Re}(A_b(\pi\pi)/A_{\Sigma}(\pi\pi))}{\mathrm{Im}(A_b(\pi\pi)/A_{\Sigma}(\pi\pi))} \,,
\end{align}
and can be obtained from the subleading, non-universal contributions to the time-dependent CP violation observable $\Delta Y_f$, where~\cite{LHCb:2021vmn} 
\begin{align}
A_{CP}(f,t) \approx a_{CP}^{\mathrm{dir}} + \Delta Y_f \frac{t}{\tau_{D^0}}\,,\label{eq:DeltaY-1}
\end{align}
and to very good precision~\cite{LHCb:2021vmn}
\begin{align}
\Delta Y_f = x\sin\phi - 
                y \left( \left|\frac{q}{p}\right| - 1  \right) +
                y a_{CP}^{\mathrm{dir}}(f)  \left(1 + \frac{x}{y}  \cot\delta_f\right)\,. \label{eq:DeltaY-2}
\end{align}
Here, $x$, $y$, $\left|q/p\right|$ and $\phi$ are the parameters of $D^0-\overline{D}^0$ mixing, see Refs.~\cite{LHCb:2021vmn, Kagan:2020vri, Grossman:2009mn, Kagan:2009gb} for details.
Rearranging Eq.~(\ref{eq:DeltaY-2}), we extract $\cot\delta_f$ from $\Delta Y_f$ as 
\begin{align}
\cot\delta_f &= \frac{y}{x}\left( \frac{\Delta Y_f - x\sin\phi + y\left(\left|q/p\right| -1 \right)}{y a_{CP}^{\mathrm{dir}}(f)  } - 1 \right) \,. \label{eq:solution-9}
\end{align}
In terms of $\cot\delta_{KK}$, $\cot\delta_{\pi\pi}$, $\Delta a_{CP}^{\mathrm{dir}}$ and $\Sigma a_{CP}^{\mathrm{dir}}$ we obtain the following expressions for $\mathrm{Re}(\tilde{p}_0)$ and $\mathrm{Re}(\tilde{p}_1)$ to order $\mathcal{O}(\varepsilon^2)$:
\begin{align}
\mathrm{Re}(\tilde{p}_0) &= \frac{1}{8 \mathrm{Im}(\lambda_b/\Sigma)} \Delta a_{CP}^{\mathrm{dir}} \left( \cot\delta_{KK} + \cot\delta_{\pi\pi}\right)\,, \label{eq:solution-10}\\
\mathrm{Re}(\tilde{p}_1) &= 
	\frac{1}{8  \mathrm{Im}(\lambda_b/\Sigma) } \Delta a_{CP}^{\mathrm{dir}} \left( \cot\delta_{\pi\pi} (R_{KK,\pi\pi} -2 ) + \cot\delta_{KK} (R_{KK,\pi\pi} + 2)\right) + \nn\\
	 &\qquad \frac{1}{4  \mathrm{Im}(\lambda_b/\Sigma) } \Sigma a_{CP}^{\mathrm{dir}}  \left( \cot\delta_{KK} + \cot\delta_{\pi\pi} \right)\,. \label{eq:solution-11}
\end{align}

\section{Numerical Results \label{sec:numerics} }

For the numerical determination of the hadronic matrix element parameters of the parametrization Eqs.~(\ref{eq:param-1})--(\ref{eq:param-4}) we employ the experimental input data in Tables~\ref{tab:exp-input-data}, \ref{tab:mixing-correlations} and apply Eqs.~(\ref{eq:solution-1})--(\ref{eq:solution-6}) and (\ref{eq:solution-9})--(\ref{eq:solution-11}).
From the branching ratio measurements we obtain the combinations
\begin{align}
R_{K\pi} 	    &= -0.08 \pm 0.01 \,, \\
R_{KK,\pi\pi}       &= 0.532\pm 0.008 \,,\\
R_{KK,\pi\pi, K\pi} &= 0.083 \pm 0.008\,.
\end{align}
From time-dependent CP violation we obtain for the strong phases 
\begin{align}
\cot\delta_{KK} &= -28^{+61}_{-126}   \,, & 
\cot\delta_{\pi\pi} &= -28^{+30}_{-36} \,, \label{eq:cotdelta} 
\end{align}
\emph{i.e.},~basically no constraint. This is understandable from the fact that the phases only contribute to 
the subleading, final-state dependent contributions of $\Delta Y_f$, and at the current precision $\Delta Y_{K^+K^-}$ and $\Delta Y_{\pi^+\pi^-}$ do not yet show a significant final-state dependence.

Note that in principle there is a further opportunity for constraining the strong phases of $\tilde{p}_{0,1}$ by extracting $\cot\delta_{KK}$ and $\cot\delta_{\pi\pi}$ from future precision determinations of the isolated mixing parameters $y_{CP}^{KK}$ and $y_{CP}^{\pi\pi}$, where these phases also appear in subleading, final-state dependent contributions, see Refs.~\cite{Pajero:2021jev, Kagan:2020vri, Grossman:2009mn, Kagan:2009gb} for details. Recently, LHCb measured the combinations $y_{CP}^{KK}-y_{CP}^{K\pi}$ and $y_{CP}^{\pi\pi}-y_{CP}^{K\pi}$~\cite{LHCb:2022gnc}. However, like $\Delta Y_{K^+K^-}$ and $\Delta Y_{\pi^+\pi^-}$, they do not yet show a significant final-state dependence. 

Our results for all parameters in Eq.~(\ref{eq:normalized-parameters}) are given in Table~\ref{tab:results}.
As a result of Eq.~(\ref{eq:cotdelta}, we have basically no information on the real parts $\mathrm{Re}(\tilde{p}_0)$ and $\mathrm{Re}(\tilde{p}_1)$. 
We will therefore not include them in the discussion any further. 

\begin{figure}[t]
\begin{center}
\includegraphics[width=0.8\textwidth]{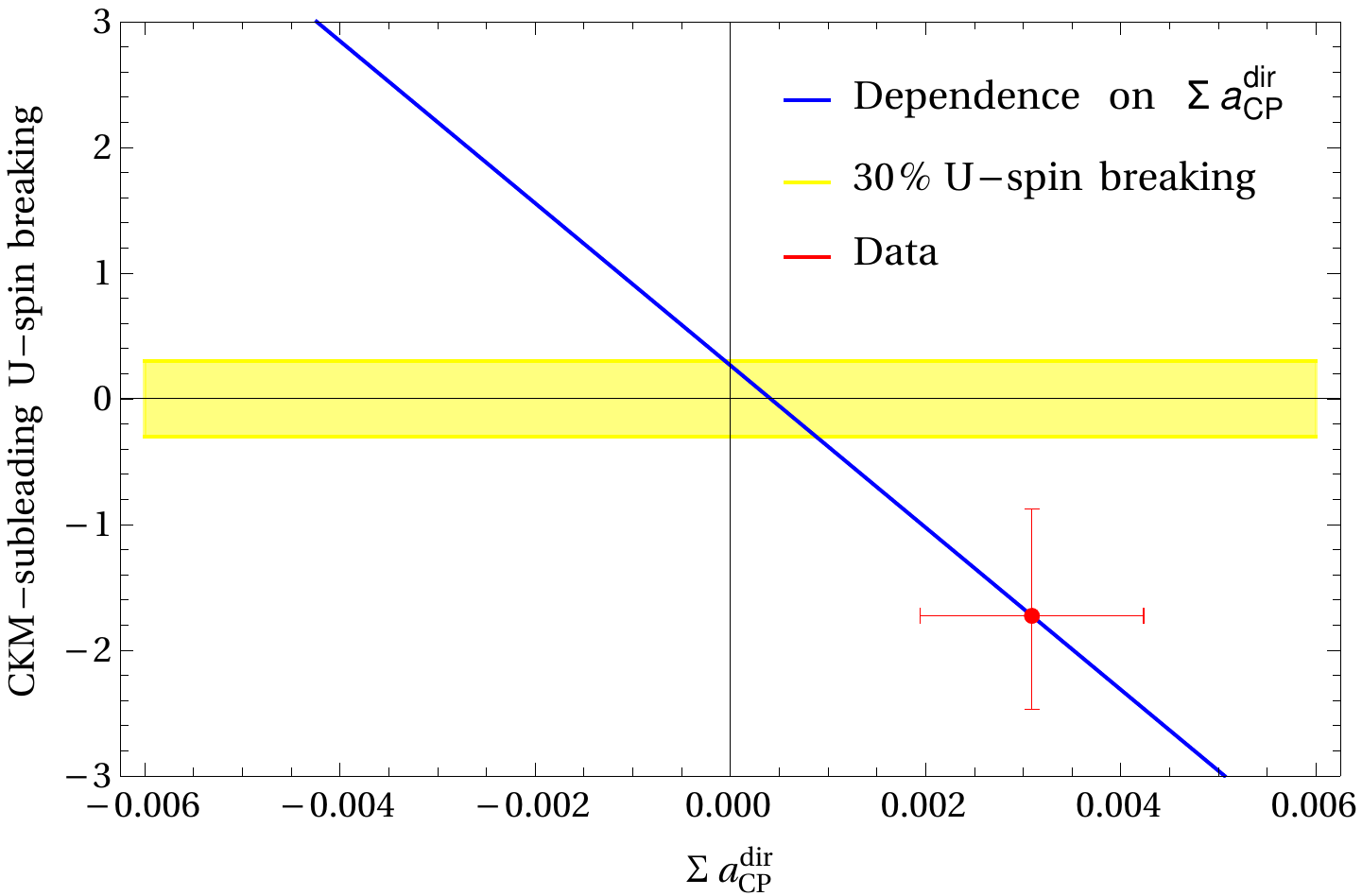}
\caption{Illustration of the dependence of $1/2\, \mathrm{Im}(\tilde{p}_1) / \mathrm{Im}(\tilde{p}_0)$ on $\Sigma a_{CP}^{\mathrm{dir}}$ according to Eq.~(\ref{eq:observable-relation}). For this illustration we fix $\Delta a_{CP}^{\mathrm{dir}}$ and $R_{KK,\pi\pi}$ to their central values and vary $\Sigma a_{CP}^{\mathrm{dir}}$ away from its measured value (blue). In red we show the current experimental data for $\Sigma a_{CP}^{\mathrm{dir}}$ and the resulting value for $1/2\, \mathrm{Im}(\tilde{p}_1) / \mathrm{Im}(\tilde{p}_0)$ including $1\sigma$ errors. 
For the estimate of the region of 30\% $U$-spin breaking (yellow) we assume that the strong phases are $\mathcal{O}(1)$, such that $ \vert 1/2\, \mathrm{Im}(\tilde{p}_1)/\mathrm{Im}(\tilde{p}_0)\vert \approx 1/2 \vert \tilde{p}_1\vert / \vert \tilde{p}_0\vert \leq 30\%$. 
\label{fig:plot}}
\end{center}
\end{figure}

We make now the following assumption: 
\begin{itemize}
\item  Due to non-perturbative rescattering~\cite{Grossman:2019xcj}, the phases of $\tilde{p}_0$ and $\tilde{p}_1$ are  $\mathcal{O}(1)$, resulting in $\left| \mathrm{Im}(\tilde{p}_1) / \mathrm{Im}(\tilde{p}_0)\right| \approx \vert \tilde{p}_1\vert/\vert \tilde{p}_0\vert$.
\end{itemize}
With future data on time-dependent CP violation this assumption can be tested and improved. 
From Eq.~(\ref{eq:observable-relation}), it follows then for the ratio of the magnitude of the $U$-spin breaking contribution to the CKM-subleading amplitude $A_b(\pi\pi)$ ($A_b(KK)$) to the corresponding $U$-spin limit contribution: 
\begin{align}
\frac{1/2\, \vert \tilde{p}_1\vert}{\vert \tilde{p}_0\vert } &\approx  \left|\frac{1/2\, \mathrm{Im}(\tilde{p}_1)}{ \mathrm{Im}(\tilde{p}_0) }\right| 
	=  1.73^{+0.85}_{-0.74}\,, \label{eq:U-spin-breaking} 
\end{align}
which deviates at $1.95\sigma$ from the SM expectation of $\mathcal{O}(30\%)$. 
Eq.~(\ref{eq:U-spin-breaking}) is our main result.
We illustrate Eq.~(\ref{eq:U-spin-breaking}) and the dependence of Eq.~(\ref{eq:observable-relation}) on $\Sigma a_{CP}^{\mathrm{dir}}$ in Fig.~\ref{fig:plot}.

The found $U$-spin breaking of $(173^{+85}_{-74})\%$ may lead to the question if the $U$-spin power counting used for its extraction in 
Eqs.~(\ref{eq:solution-1}--\ref{eq:solution-6}, \ref{eq:solution-10}, \ref{eq:solution-11}) 
is actually still valid. Note that  
Eqs.~(\ref{eq:solution-1}--\ref{eq:solution-6}, \ref{eq:solution-10}, \ref{eq:solution-11}) 
are all formally valid at $\mathcal{O}(\varepsilon^2)$. 
Now, if $\mathrm{Im}(\tilde{p}_1)$ breaks the power counting by being $\mathcal{O}(1)$ instead of $\mathcal{O}(\varepsilon)$, these equations have the following power counting: 
\begin{itemize}
\item Eqs.~(\ref{eq:solution-1}--\ref{eq:solution-4}, \ref{eq:solution-6}) for the extraction of  
$\mathrm{Re}(\tilde{t}_1)$, $\mathrm{Im}(\tilde{t}_1)$, $\tilde{t}_2$, $\tilde{s}_1$, and 
$\mathrm{Im}(\tilde{p}_1)$ are in this case still valid at $\mathcal{O}(\varepsilon^2)$.
\item Eqs.~(\ref{eq:solution-5}, \ref{eq:solution-11}) for the extraction of $\mathrm{Im}(\tilde{p}_0)$ and $\mathrm{Re}(\tilde{p}_1)$ are in this case valid at $\mathcal{O}(\varepsilon)$.
\item Eq.~(\ref{eq:solution-10}) obtains $\mathcal{O}(1)$ corrections, \emph{i.e.}~is broken in this case and can no longer be used for the extraction of  $\mathrm{Re}(\tilde{p}_0)$.
\end{itemize}
Note that also Eq.~(\ref{eq:observable-relation}) is formally valid at $\mathcal{O}(\varepsilon^2)$ and is still valid at $\mathcal{O}(\varepsilon)$ when $\mathrm{Im}(\tilde{p}_1)\sim\mathcal{O}(1)$. 
The above implies that for $\mathrm{Im}(\tilde{p}_1)\sim \mathcal{O}(1)$ the methodology of Sec.~\ref{sec:analytic} still enables a consistent parameter extraction with the exception of the parameter $\mathrm{Re}(\tilde{p}_0)$. However, with current data we have in any case no sensitivity to this parameter. 
As can be seen from Table~\ref{tab:results}, the other $U$-spin breaking parameters are consistent with the $U$-spin power counting.

\begin{table}[t]
\centering
\begin{tabular}{|c|c|c|}
\hline\hline
\multicolumn{3}{|c|}{Direct CP Asymmetries}\\
\hline

$a_{CP}^{\mathrm{dir}}(D^0\rightarrow K^+K^-)$ & $(7.7\pm 5.7)\cdot 10^{-4} $ &  \cite{Maccolini:2022}\\

$a_{CP}^{\mathrm{dir}}(D^0\rightarrow\pi^+\pi^-)$ & $(23.2\pm 6.1)\cdot 10^{-4}$ & \cite{Maccolini:2022} \\

$\rho$ & 0.88 & \cite{Maccolini:2022} \\\hline

\multicolumn{3}{|c|}{Time-dependent CP Violation}\\\hline

$\Delta Y_{K^+K^-}$     &  $(-2.3\pm 1.5\pm 0.3)\cdot 10^{-4}$  & \cite{LHCb:2021vmn} \\ 
$\Delta Y_{\pi^+\pi^-}$ &  $(-4.0\pm 2.8\pm 0.4)\cdot 10^{-4}$  & \cite{LHCb:2021vmn} \\\hline

\multicolumn{3}{|c|}{$D^0-\overline{D}^0$ Mixing Parameters}\\\hline
 
$x$     	 & $\left(0.409_{-0.049}^{+0.048}\right)\cdot 10^{-2}$  & \cite{HFLAV:2022pwe}  \\  
$y$     	 & $\left(0.615^{+0.056}_{-0.055}\right)\cdot 10^{-2} $   & \cite{HFLAV:2022pwe}   \\ 
$\delta_{K\pi}$  & $(7.2^{+7.9}_{-9.2})^{\circ}$ & \cite{HFLAV:2022pwe} \\ 
$\vert q/p\vert$ &  $0.995\pm 0.016$ & \cite{HFLAV:2022pwe}  \\
$\phi$  	 &  $\left(-2.5\pm 1.2\right)^{\circ}$  & \cite{HFLAV:2022pwe}  \\\hline
 
\multicolumn{3}{|c|}{Branching Ratios}\\
\hline

$\mathcal{B}(D^0\rightarrow K^+K^-)$    & $(4.08\pm 0.06)\cdot 10^{-3}$  & \cite{Workman:2022ynf}\\

$\mathcal{B}(D^0\rightarrow\pi^+\pi^-)$ & $(1.454\pm 0.024)\cdot 10^{-3}$   & \cite{Workman:2022ynf}\\

$\mathcal{B}(D^0\rightarrow K^+\pi^-)$  & $(1.363\pm 0.025)\cdot 10^{-4}$  & \cite{Workman:2022ynf}\\

$\mathcal{B}(D^0\rightarrow K^-\pi^+)$  & $(3.947\pm 0.030)\cdot 10^{-2}$  &  \cite{Workman:2022ynf} \\\hline

\multicolumn{3}{|c|}{Further Numerical Inputs}\\
\hline

$\mathrm{Im}\left(\lambda_b/\Sigma\right)$ & $(-6.0 \pm 0.3)\cdot 10^{-4}$  & \cite{Workman:2022ynf} \\\hline\hline 

\end{tabular}
\caption{Experimental input data. We include the correlation between $a_{CP}^{\mathrm{dir}}(D^0\rightarrow K^+K^-)$ and $a_{CP}^{\mathrm{dir}}(D^0\rightarrow \pi^+\pi^-)$ which is given by $\rho$. 
We also include the correlations between the mixing parameters $x$, $y$, $\delta_{K\pi}$, $\vert q/p\vert$, and $\phi$, which are given in Table~\ref{tab:mixing-correlations}. 
We symmetrize all errors of the input data if applicable. 
Note that for $a_{CP}^{\mathrm{dir}}(D^0\rightarrow K^+K^-)$ and $a_{CP}^{\mathrm{dir}}(D^0\rightarrow\pi^+\pi^-)$ we use directly the most recent preliminary measurements by LHCb which contain all Run-1 and Run-2 measurements.
Note further that the fit results by the Heavy Flavor Averaging Group (HFLAV) for the $D^0-\overline{D}^0$ mixing parameters \cite{HFLAV:2022pwe} do not yet include these latest results for the direct CP asymmetries. 
Both of these points can be improved in the future with updates of the world averages and global fits~\cite{HFLAV:2022pwe, BaBar:2007tfw, CDF:2012ous, LHCb:2014kcb, LHCb:2016csn,Belle:2008ddg, Belle:2015etc}. 
\label{tab:exp-input-data}}
\end{table}

\begin{table}[t]
\centering
\begin{tabular}{|c|c|c|c|c|c|}
\hline\hline

        & $x$  & $y$   & $\delta_{K\pi}$ & $\vert q/p\vert$ & $\phi$  \\\hline

$x$		& $1.0$ & $-0.075$ & $-0.029$  & $-0.122$ & $0.087$  \\\hline 
$y$     	& $-0.075$ & $1.0$ & $0.970$ & $-0.035$  & $0.071$\\\hline    
$\delta_{K\pi}$ & $-0.029$  & $0.970$ & $1.0$ &  $-0.043$  &  $0.079$  \\\hline    
$\vert q/p\vert$& $-0.122$  & $-0.035$   & $-0.043$   & $1.0$& $0.558$ \\\hline    
$\phi$  	& $ 0.087$  & $0.071$  & $0.079$  & $0.558$  & $1.0$ \\   \hline\hline 

\end{tabular}
\caption{Correlation matrix for the needed $D^0-\overline{D}^0$ mixing parameters from Ref.~\cite{HFLAV:2022pwe}.  
\label{tab:mixing-correlations}}
\end{table}

\begin{table}[t]
\centering
\begin{tabular}{|c|c|}
\hline\hline
$\mathrm{Re}(\tilde{t}_1)$  & $0.083\pm 0.010$ \\
$\mathrm{Im}(\tilde{t}_1)$  & $-0.11^{+0.15}_{-0.16}$  \\
$\tilde{t}_2$ &  $0.101^{+0.019}_{-0.016}$ \\  
$\tilde{s}_1$ & $-0.2658^{+0.0040}_{-0.0039}$   \\ 
$\mathrm{Im}(\tilde{p}_0)$ &  $0.66 \pm 0.13$  \\
$\mathrm{Im}(\tilde{p}_1)$ & $-2.27^{+0.96}_{-0.98}$  \\
$\mathrm{Re}(\tilde{p}_0)$ & $-18^{+23}_{-47}$ \\
$\mathrm{Re}(\tilde{p}_1)$ & $64^{+56}_{-55}$  \\\hline\hline
\end{tabular}
\caption{Results for hadronic matrix elements of the $U$-spin expansion Eqs.~(\ref{eq:param-1})--(\ref{eq:param-4}), as extracted from the experimental data in Tables~\ref{tab:exp-input-data} and \ref{tab:mixing-correlations}. 
\label{tab:results}}
\end{table}

\clearpage

\section{Predictions and New Physics Interpretations \label{sec:predictions}}

The large $U$-spin breaking of $(173^{+85}_{-74})\%$ that we find in Eq.~(\ref{eq:U-spin-breaking}) indicates large contributions from  
$\Delta U=1$ operators in the CKM-subleading amplitude of SCS charm decays. 
This leads to an $\mathcal{O}(1)$ breaking of the $U$-spin limit sum rule~\cite{Grossman:2006jg, Pirtskhalava:2011va, Hiller:2012xm, Grossman:2012ry}
\begin{align}
\frac{\Gamma(D^0\rightarrow K^+K^-)}{\Gamma(D^0\rightarrow \pi^+\pi^-)} &= -\frac{a_{CP}^{\mathrm{dir}}(D^0\rightarrow \pi^+\pi^-)}{a_{CP}^{\mathrm{dir}}(D^0\rightarrow K^+K^-)}\,, \label{eq:sum-rule-1} 
\end{align}
see the discussion in Sec.~\ref{sec:introduction}. 
As the decays $D^0\rightarrow K^+K^-$ and $D^0\rightarrow \pi^+\pi^-$ are also connected to a wider class of decays via   
SU(3)$_F$ symmetry, we expect that the $U$-spin limit sum rule~\cite{Grossman:2006jg, Pirtskhalava:2011va, Hiller:2012xm, Grossman:2012ry}
\begin{align}
\frac{\Gamma(D^+\rightarrow \overline{K}^0 K^+)}{\Gamma(D_s^+\rightarrow K^0\pi^+)} &= - \frac{a_{CP}^{\mathrm{dir}}(D_s^+\rightarrow K^0\pi^+)}{a_{CP}^{\mathrm{dir}}(D^+\rightarrow \overline{K}^0 K^+)} \label{eq:sum-rule-2}
\end{align}
is also broken at $\mathcal{O}(1)$. In Ref.~\cite{Muller:2015rna} improved versions of the sum rules Eqs.~(\ref{eq:sum-rule-1}, \ref{eq:sum-rule-2}) are formulated that account for the first order SU(3)$_F$ breaking effects from all topological diagrams except for the penguin contraction of tree operators ($P$ and $PA$ in the notation therein). Therefore, we predict that also the sum rules~\cite{Muller:2015rna}
\begin{align}
\frac{\mathcal{S}(D^0\rightarrow K^+K^-) - \mathcal{S}(D^0\rightarrow \pi^+\pi^-)}{e^{2i\delta(D^0\rightarrow K^+K^-)} - e^{2i\delta(D^0\rightarrow\pi^+\pi^-)}} -
 \frac{\mathcal{S}(D^0\rightarrow K^+K^-) + \sqrt{2} \mathcal{S}(D^0\rightarrow \pi^0\pi^0)}{e^{2i\delta(D^0\rightarrow K^+K^-)} - e^{2i\delta(D^0\rightarrow\pi^0\pi^0)}} &= 0\,, \label{eq:improved-sum-rule-1}\\
\frac{\mathcal{S}(D^+\rightarrow \overline{K}^0 K^+) - \mathcal{S}(D_s^+\rightarrow K^0\pi^+)}{e^{2i\delta(D^+\rightarrow \overline{K}^0 K^+)} - e^{2i\delta(D_s^+\rightarrow K^0 \pi^+)}} -
 \frac{\mathcal{S}(D^+\rightarrow \overline{K}^0 K^+) + \sqrt{2} \mathcal{S}(D_s^+\rightarrow K^+\pi^0)}{e^{2i\delta(D^+\rightarrow \overline{K}^0 K^+)} - e^{2i\delta(D_s^+ \rightarrow K^+ \pi^0 )}} &= 0  \label{eq:improved-sum-rule-2}
\end{align}
are broken at $\mathcal{O}(1)$. Here, $\delta(d) \equiv \mathrm{arg}(A_{\Sigma}(d))$ and the function $\mathcal{S}(d)$ can be found in Ref.~\cite{Muller:2015rna}. 
Further SU(3)$_F$ sum rules are given in Refs.~\cite{Grossman:2012ry, Grossman:2013lya}. For a general treatment of $U$-spin sum rules at any order see Ref.~\cite{Gavrilova:2022hbx}. 
In light of the puzzle posed by the $U$-spin expansion of charm decays, also a further test of the respective isospin structure is very important~\cite{Grossman:2012eb}.

As laid out in Ref.~\cite{Hiller:2012xm}, new physics models with additional $\Delta U=1$ operators, so-called \lq\lq{}$\Delta U=1$ models\rq\rq{}~\cite{Hiller:2012xm} can explain the breaking of Eqs.~(\ref{eq:sum-rule-1}, \ref{eq:sum-rule-2}) beyond the $U$-spin power counting. The same applies to Eqs.~(\ref{eq:improved-sum-rule-1}, \ref{eq:improved-sum-rule-2}).
Such models generate additional effective operators with the flavor content $\bar{s}c\bar{u}s$ and/or $\bar{d}c\bar{u}d$ with non-universal coefficients. They can \emph{e.g.}~arise from two-Higgs-doublet models (2HDMs) \cite{Altmannshofer:2012ur} or flavorful $Z'$ models \cite{Altmannshofer:2012ur, Chala:2019fdb, Bause:2020obd}. 
Recently, in Ref.~\cite{Bause:2020obd} it has been shown that $Z'$ models can induce large $U$-spin breaking between $a_{CP}^{\mathrm{dir}}(D^0\rightarrow K^+K^-)$ and $a_{CP}^{\mathrm{dir}}(D^0\rightarrow \pi^+\pi^-)$, depending on the charge assignments of the quarks under an additional $U(1)'$ group. 
With the new data, charm CP asymmetries can be used effectively to probe and explore the parameter 
space of such models further.
For example, the specific charge assignments of $Z'$ models considered in Ref.~\cite{Bause:2020obd} lead to opposite signs for $a_{CP}^{\mathrm{dir}}(D^0\rightarrow K^+K^-)$ and $a_{CP}^{\mathrm{dir}}(D^0\rightarrow \pi^+\pi^-)$, see Fig.~3 therein, whereas the most recent data indicates $a_{CP}^{\mathrm{dir}}(D^0\rightarrow K^+K^-) > 0$ and $a_{CP}^{\mathrm{dir}}(D^0\rightarrow \pi^+\pi^-) >0$.

The exploration of the $U$-spin puzzle with future and more precise measurements including sum rule tests is important for a complete understanding of the CKM-subleading amplitudes of SCS charm decays and in order to further probe the parameter space of $\Delta U=1$ models.

\section{Conclusions \label{sec:conclusions}}

Assuming the Standard Model, from recent measurements of charm CP violation in the single decay channels $D^0\rightarrow K^+K^-$ and $D^0\rightarrow \pi^+\pi^-$ we 
extract for the first time the imaginary part $\mathrm{Im}(\tilde{p}_1)$ of the $U$-spin breaking $\Delta U=1$ contribution to the CKM-subleading amplitudes. We obtain 
\begin{align}
\frac{1/2 \, \mathrm{Im}(\tilde{p}_1)}{\mathrm{Im}(\tilde{p}_0)} &= (-173^{+74}_{-85})\% \,,
\end{align}
where $\mathrm{Im}(\tilde{p}_0)$ is the $U$-spin limit $\Delta U=0$ contribution to the CKM-subleading amplitudes which is determined by $\Delta a_{CP}^{\mathrm{dir}}$. The strong phases of $\tilde{p}_{0,1}$ are yet unknown. 
Assuming $\mathcal{O}(1)$ strong phases due to non-perturbative rescattering, the result implies very large $U$-spin breaking, which exceeds the SM expectation of $\sim 30\%$ by almost a factor six, at $1.95\sigma$. 

It is crucial to probe this anomaly further with more data and test the $U$-spin expansion also in additional decays, using the sum rules listed in Sec.~\ref{sec:predictions}. Most importantly, we need improved time-dependent measurements, such that we can extract the strong phases of $\tilde{p}_{0,1}$ from data. In order to test the pattern of the SU(3)$_F$ expansion, measurements of CP asymmetries of basically all singly-Cabibbo suppressed decays are necessary. 

We encourage experimental collaborations to extract the underlying theory parameters using the methodology 
described in Sec.~\ref{sec:analytic} directly from the data, enabling the most comprehensive treatment of 
all correlations. 

If the $U$-spin anomaly is confirmed with more data in the future, this would imply either a breakdown of the $U$-spin expansion in the Standard Model, or a sign for new physics with an additional $\Delta U=1$ operator, for example from additional scalar particles or a flavorful~$Z'$.

\begin{acknowledgements}
We thank Laurent Dufour, Marco Gersabeck, Yuval Grossman, Ulrich Nierste, Guillaume Pietrzyk and Alan Schwartz for useful discussions.
S.S. is supported by a Stephen Hawking Fellowship from UKRI under reference EP/T01623X/1 and the Lancaster-Manchester-Sheffield Consortium for Fundamental Physics, under STFC research grant ST/T001038/1.
For the purpose of open access, the author has applied a Creative Commons Attribution (CC BY) licence to any Author Accepted Manuscript version arising.
This work uses existing data which is available at locations cited in the bibliography.
\end{acknowledgements}


\bibliography{draft.bib}

\bibliographystyle{apsrev4-1}

\end{document}